# Far Infrared Sensing Of The Oxidation State Of Heme Proteins

J.-Y. Chen, J. Cerne and A. G. Markelz

*Physics Department, University at Buffalo*

## Abstract

We propose a biosensor based on the change of terahertz (THz) dielectric response when a biotarget binds to a probe. The feasibility of this biosensor is examined by studying the change of THz transmission for thin films of heme proteins myoglobin (Mb) and cytochrome C (CytC) as a function oxygen binding. We measure a strong increase in both the index and absorbance with oxygen binding for both Mb and CytC. The measured changes occur for both dried and samples hydrated at 80% r.h. suggesting that using terahertz time domain spectroscopy (TTDS) to monitor the transmitted terahertz pulse through a biomolecular probe thin film in ambient environmental conditions could be used to determine the presence of a biomolecular target.

## I. Introduction

Many current biosensor techniques rely on engineered probes to sense target molecule binding.[1,2] A simpler method is to monitor a physical characteristic that must change with binding such as mass. For example, in microcantilever biosensors the biorecognition element is attached to one side of the cantilever.[3] Upon binding of the target to the biorecognition element, the mass loading causes cantilever deflection monitored using standard atomic force microscope techniques. This technique has shortcomings: 1.) sophisticated processing of the cantilever tips required; 2.) complicated position sensitive detection; and 3.) vulnerability to turbulence and vibrations. The biomolecular vibrational modes involving collective movement of protein structural domains lay in the THz range.[4] Binding of the target will impact the amplitude and oscillator strength of these modes: for example they may red shift due to by mass loading, reduce their amplitude due to steric constraints or diminish/increase in oscillator strength due to electronic redistribution with binding. Thus by monitoring the THz absorbance of a biomolecular probe molecule, one can determine the presence of a bound target by the change in THz absorbance is the system is sufficiently sensitive and the change of THz absorbance is sufficiently strong. In this paper we examine the change in THz absorbance with simple ligand binding to biomolecules, that is the oxidation of heme proteins myoglobin and cytochrome c.

## II. Materials and Methods

TTDS is performed on thin films of cytochrome c and myoglobin on infrasil quartz. The oxidation state is prepared in solutions from which the films are formed. Lyophilized powders of myoglobin (sigma no. M-0630) and cytochrome C (sigma no. C-2037) were purchased from Sigma and used without further purification. High concentration solutions were made by dissolving 40 mg protein in 200 µl of Tris buffer (ph 7.0). Metmyoglobin/oxycytochrome C were converted into deoxy-myoglobin/deoxy-cytochrome C by adding excess 20 mg/ml sodium dithionite into metmyoglobin/oxycytochrome c solutions. Thin films were formed by pipetting 10 μl of the protein solution on infrasil quartz substrates. One half of the substrate was left clean to be used as a reference. The samples were placed in a box purged by nitrogen gas for ~ 8 h. to form nonflowing uniform thin films. To achieve sufficient optical density, the pipetting and drying procedure was repeated twice. Typical thicknesses were ~ 200 +/- 20 μm. The oxidation state of the films was verified using UV/Vis absorption measurements.[5-7]

A standard THz time domain spectroscopy system is used to monitor the change in absorbance and index with oxidation state.[8] The THz is generated using a hertzian dipole antenna and detected using electro optic detection or antenna detection using a photoconductive switch on ion implanted silicon on sapphire.[9, 10] Samples are mounted in a humidity controlled cell.[11] Two hydrations are studied here, dry (< 5% r.h.) and hydrated, (80 % r.h.). The hydration cell is flushed with either dry gas or gas hydrated by flowing over a saturated salt solution. For oxidized samples the flow gas is oxygen, for the reduced samples the flow gas is air. The typical time for flushing before data taking was >2 h. By performing the measurements as a function of hydration we can get some idea of the role water plays in the change in the THz response. All measurements were performed at room temperature. TTDS which measures the transmitted *field*, allows us to both access the real and imaginary part of the dielectric response in a single measurement. Thus we can both see how the index and the absorbance change as a function of frequency. The field transmission is determined by the measured transmitted field for the sample and reference through:

$$t = \frac{E_{sample}(\omega)e^{i\phi_{sample}(\omega)}}{E_{ref}(\omega)e^{i\phi_{ref}(\omega)}} = |t|e^{i\phi_t(\omega)} \quad (1)$$

Where $\phi_t(\omega)$ is the phase. The absorbance is then given by:

$$A(\omega) = -2\log(|t|) \quad (2)$$

### III. Results

In Figure 1 we show the phase of the field transmission for the myoglobin films. As seen in the figure, one can readily identify the metMb film from the deoxyMb film through the THz phase. The nominal index of the metMb for purged and 80% relative humidity are 1.48 and 1.60 respectively. The nominal indices for the deoxyMb purged and at 80% r.h. are 1.27 and 1.33 respectively. The overall smaller indices for the deoxy samples suggest a net decrease in polarizability, consistent with the ionic nature of the oxygen bonding. The absorbance measured for the myoglobin films is shown in Figure 2. The absorbance is higher for the oxidized samples. Calculations of the normal modes of the met and deoxy myoglobin are nearly identical, suggesting that the increase in the absorbance is due to the increase in dipole moment with oxidation. For both met and deoxy samples the absorbance increases with water content. This increase may be due to either an increase in flexibility of the molecules with water mediated internal bonding or the increased absorbance may arise from the absorbed water itself absorbing THz.

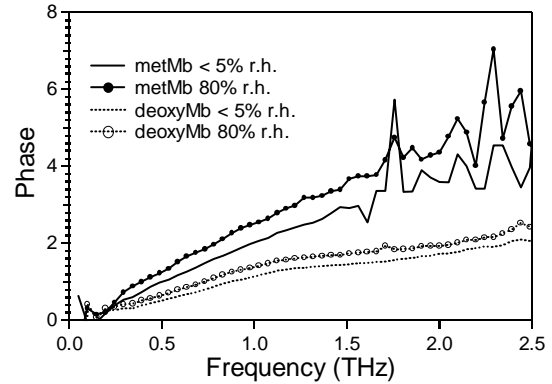

**Figure 1. Phase of the field transmission for metMb and deoxyMb thin films. See Eq. 1.**

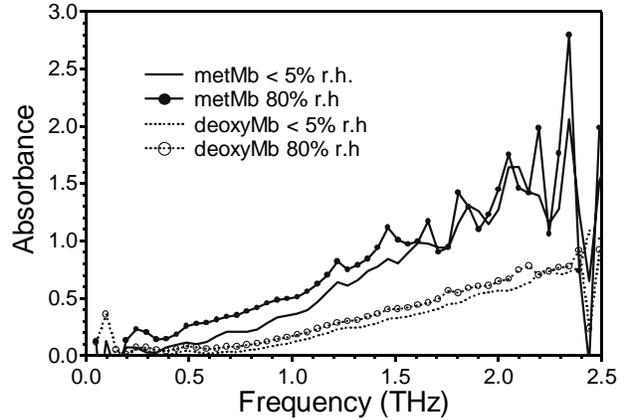

Figure 2. Terahertz absorbance for metMb and deoxyMb thin films. See Eq. 2.

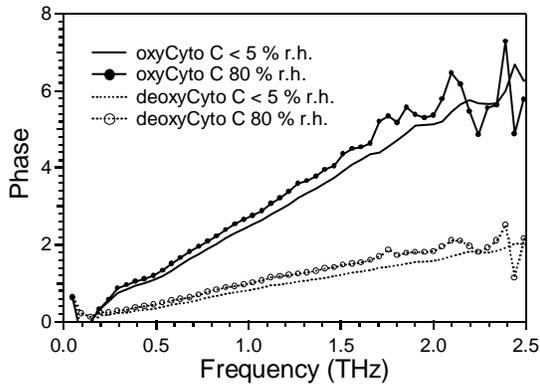

Figure 3. Phase of the field transmission for oxyCytC and deoxyCytC thin films. See Eq. 1.

The phase and absorbance measurements for the cytochrome C samples are shown in Figures 3 and 4 respectively. Again the phase and absorbance are clear indicators of the oxidation state of the biomolecular samples. The nominal indices determined from the phase measurements for purged (80% r.h.) samples are 1.59 (1.64) for oxyCyto C, and 1.19 (1.24) for deoxyCyto C. Again the oxidized biomolecules have a significantly higher index, indicating higher polarizability. The difference in the absorbance is more dramatic in the cytochrome C case. The index and absorbance increase with hydration for all samples, but less dramatically than in the myoglobin case.

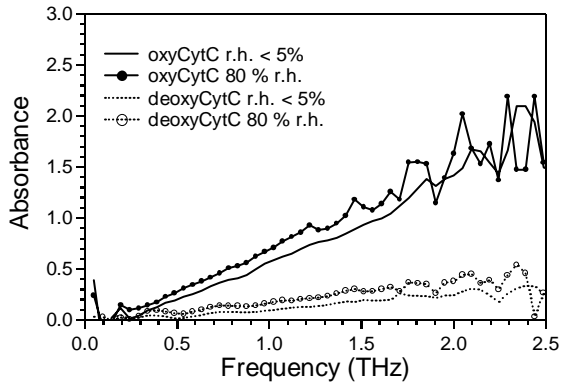

Figure 4. Terahertz absorbance for oxyCytC and deoxyCytC thin films. See Eq. 2.

### IV. Conclusions

We find a significant difference in the both the real and imaginary parts of the THz dielectric response as a function of oxidation for myoglobin and cytochrome C samples. The phase sensitivity to ligand binding is interesting in view of sensor fabrication in that the presence of the target can be immediately determined by the change in arrival time of the transmitted pulse. This result is similar to Haring-Bolivar and co workers who showed high sensitivity to DNA binding in dielectric index measurements using a stripline transmission approach.[12] A number of calculations and followup measurements are suggested by the results as well as future measurements. Among the issues raised in these studies are: the dependence of THz response for large ligand binding more akin to actual biosensor applications, such as antigen-antibody binding; predicted polarizability changes determined by molecular modeling; and a detailed understanding of the absorbed water and threshold for liquid like water behavior.

**Acknowledgements**

We gratefully acknowledge the funding of this work through Army Research Office grant DAAD19-02-1-0271.